\documentclass[sigconf]{acmart}

\AtBeginDocument{%
  \providecommand\BibTeX{{%
    \normalfont B\kern-0.5em{\scshape i\kern-0.25em b}\kern-0.8em\TeX}}}

\copyrightyear{2026}
\acmYear{2026}
\setcopyright{cc}
\setcctype{by}
\acmConference[IGSC 2026]{International Green and Sustainable Computing Conference}{June 22--24, 2026}{Canandaigua, NY, USA}
\acmBooktitle{International Green and Sustainable Computing Conference (IGSC 2026), June 22--24, 2026, Canandaigua, NY, USA}
\acmDOI{10.1145/3797248.3815400}
\acmISBN{979-8-4007-2520-3/2026/06}

\begin{document}

\title{CarbonSim: A Lifecycle-Aware Framework for Evaluating Carbon Tradeoffs in Hardware Upgrade Decisions}
\author{Kartik Hans}
\affiliation{%
  \department{Department of Computer Science}
  \institution{University of Pittsburgh}
  \city{Pittsburgh}
  \state{PA}
  \country{USA}
}
\email{hans@sci.pitt.edu}

\author{Kaiwen Zhao}
\affiliation{%
  \department{Department of Computer Science}
  \institution{University of Pittsburgh}
  \city{Pittsburgh}
  \state{PA}
  \country{USA}
}
\email{kaz81@pitt.edu}

\author{Stephen Lee}
\affiliation{%
  \department{Department of Computer Science}
  \institution{University of Pittsburgh}
  \city{Pittsburgh}
  \state{PA}
  \country{USA}
}
\email{stephen.lee@pitt.edu}

\begin{abstract}
As the demand for information and communication technologies (ICT) continues to rise, the environmental impact of computing systems is becoming an increasingly critical concern. Although newer hardware often improves performance and energy efficiency, these gains do not always offset the carbon cost of premature replacement, particularly under low-utilization workloads or low-carbon electricity grids.
We present CarbonSim, a lifecycle-aware simulation framework for evaluating carbon tradeoffs in hardware upgrade decisions. CarbonSim combines workload execution profiles, machine-level power characteristics, embodied carbon inventories, scheduling policies, and time-varying grid carbon intensity to estimate total emissions under alternative deployment scenarios. The framework supports multiple embodied-carbon accounting strategies, including uniform amortization and front-loaded lifecycle attribution, enabling analysis under different hardware lifespan assumptions.
Using heterogeneous CPU generations as calibration platforms, we demonstrate that newer machines do not always minimize total emissions: under lightly loaded workloads or cleaner electricity mixes, extending the useful life of existing hardware can reduce lifecycle carbon despite lower operational efficiency. These results highlight that hardware refresh decisions should be workload-aware, location-aware, and lifecycle-aware. 
\end{abstract}

\ccsdesc[500]{Computer systems organization~Cloud computing}
\ccsdesc[500]{Hardware~Impact on the environment}
\ccsdesc[300]{Computer systems organization~Availability}

\keywords{carbon emissions, embodied carbon, hardware lifecycle, carbon-aware scheduling, sustainability}

\maketitle

\section{Introduction}

The environmental impact of computing systems is increasing as the demand for information and computing technologies (ICT) continues to grow. Recent studies indicate that ICT-related operational emissions could account for up to 8\% of global greenhouse gas emissions within the next decade if left unmitigated~\cite{andrae2015global}. This rise is driven by energy-intensive activities in data centers, networking infrastructure, and end-user devices, compounded by the rapid growth of AI, cloud computing, and the Internet of Things (IoT). Additionally, embodied emissions --- those arising from the manufacturing, transportation, and disposal of ICT hardware --- are becoming a significant concern~\cite{acun2023carbon, gupta2022act}. As such, recent studies have highlighted the importance of addressing these emissions by extending the operational lifetime of devices~\cite{chien2021driving, patterson2021carbon, lechowicz2024carbonclipper, deng2012policy, gupta2021chasing,switzer2023junkyard}. In modern computing systems, these embodied costs can affect whether replacing existing hardware yields a net environmental benefit. 

Historically, frequent hardware replacement was often justified by the strong performance-per-watt gains associated with successive technology generations. Under earlier phases of Moore’s Law and Dennard scaling, newer systems typically delivered sufficiently large efficiency improvements to offset the operational cost of maintaining older infrastructure~\cite{serverefficiency}. However, as transistor scaling slows and efficiency gains become less pronounced, the environmental case for replacement becomes less obvious ~\cite{prieto2025evolution}. Although newer platforms often execute workloads faster, their embodied carbon cost may outweigh operational savings when utilization is low, workloads are bursty, or electricity is already low carbon.

Evaluating this tradeoff is complicated because upgrade decisions operate across multiple interacting timescales: workload behavior changes at runtime, electricity carbon intensity varies hourly, and embodied emissions are amortized over multi-year lifecycles. In practice, datacenter-wide carbon outcomes also depend on cooling systems, memory, storage, accelerators, and facility overhead. In this work, we intentionally isolate compute-node effects as a first step and model facility overhead as an extensible parameter rather than a fixed assumption. This enables controlled analysis of how workload placement and hardware age jointly affect lifecycle emissions before incorporating full datacenter complexity. This motivates a central systems question: under what workload conditions, electricity mixes, and hardware lifetimes does replacing existing compute hardware reduce total lifecycle carbon emissions?

To address this challenge, we present CarbonSim, a lifecycle-aware simulation framework for evaluating carbon tradeoffs in hardware upgrade decisions across heterogeneous compute fleets. CarbonSim combines workload execution profiles, machine-specific power characteristics, embodied carbon inventories, and time-varying electricity carbon intensity to estimate total emissions under alternative scheduling and deployment strategies. 
Unlike prior work~\cite{raith2023faas, bashir2024promise, xu2018renewable,wen2023k8ssim, calheiros2011cloudsim}, our approach incorporates workload and server utilization characteristics to inform an accounting model that integrates both embodied and operational emissions.
Existing frameworks such as ACT emphasize embodied estimation at component design time~\cite{gupta2022act}, while cloud simulators such as CloudSim~\cite{calheiros2011cloudsim} and K8sSim~\cite{wen2023k8ssim} focus primarily on operational scheduling, but few tools allow these two dimensions to be jointly explored when evaluating hardware retention or replacement decisions.

Our contributions are as follows. (1) We introduce CarbonSim, an extensible simulation framework that jointly models operational and embodied carbon for hardware upgrade analysis under heterogeneous workload and scheduling conditions. (2) We formalize two embodied-carbon accounting strategies—uniform amortization and front-loaded lifecycle attribution—to study how lifespan assumptions alter upgrade outcomes.
(3) Using cross-generation calibration platforms, we show that newer hardware does not always minimize lifecycle emissions: under light workloads or cleaner grids, extending hardware lifetime can reduce total carbon.
(4) We demonstrate that mixed-generation clusters under carbon-aware scheduling can reduce emissions, while revealing explicit performance tradeoffs that future schedulers must manage. The
\texttt{CarbonSim} code is available\footnote{https://github.com/pittcps/carbonsim}.

\section{Case for Aging Machines}
Hardware energy efficiency has historically improved in tandem with Moore's Law. This trend has enabled performance-per-watt improvements over time, driven by technological advancements including architectural optimizations and better power management techniques. However, the gains in energy efficiency have not scaled linearly with newer generations. Physical limitations and a slow down of Moore's Law has led to diminishing returns in energy proportionality. Recent studies suggest that while energy efficiency will continue to improve in future servers, the rate of improvement is expected to slow considerably and may plateau in the coming years~\cite{serverefficiency}. As a result, replacing older machines does not automatically guarantee proportional reductions in lifecycle carbon emissions.

In our work, we use heterogeneous compute platforms spanning multiple CPU generations as controlled calibration points to examine generational trends in execution time, power use, and embodied-carbon sensitivity. These systems are not intended to represent full server deployments or datacenter-wide measurements. Instead, they provide reproducible cross-generation observations that help isolate how hardware age interacts with workload intensity and carbon accounting assumptions. CarbonSim is designed so that server-grade hardware profiles, benchmark traces, and facility-level parameters can be substituted directly through configuration.

Newer compute platforms often incorporate higher core counts and larger caches to improve peak performance\cite{serverefficiency1}. While these features increase throughput, they can also raise idle or lightly loaded power consumption. For workloads that do not fully utilize available resources, the marginal operational gains of newer hardware may therefore be limited, especially when embodied emissions associated with manufacturing are considered. Under such conditions, extending the service life of existing machines may remain environmentally favorable despite lower peak performance.


\begin{figure}[t]
    \centering
    \includegraphics[width=2.8in]{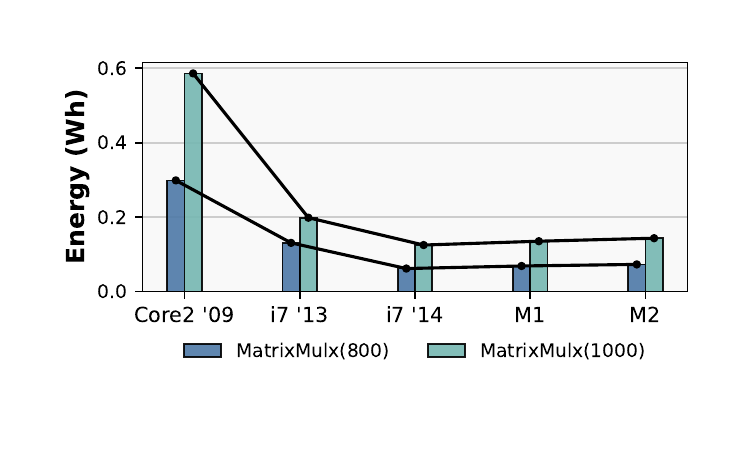}
    \vspace{-0.3in}
    \caption{Overall energy consumption for different machines.}
    \label{fig:energy}
\end{figure}
{\bf Energy Consumption Comparison.} To examine this effect, we executed representative workloads on compute platforms spanning multiple CPU generations and analyzed their energy consumption characteristics. Figure~\ref{fig:energy} illustrates the total energy consumption of running a matrix multiplication workload across these different CPU generations. As shown, older machines such as those from 2009 exhibit significantly higher energy usage. However, we observe that a 2013/2014-era server consumes a similar amount of energy to a 2020/22 model when executing the same workload. This suggests that, for certain types of computation, the energy efficiency improvements between mid-generation and modern servers may be marginal ---particularly when not fully utilizing the capabilities of newer hardware. \textit{These findings highlight that newer machines do not always result in lower energy consumption for certain workloads. In scenarios where the workload does not fully utilize the available resources, the higher idle power of newer servers can lead to greater overall energy use.}

\begin{figure}[t]
    \centering
    \includegraphics[width=2.8in]{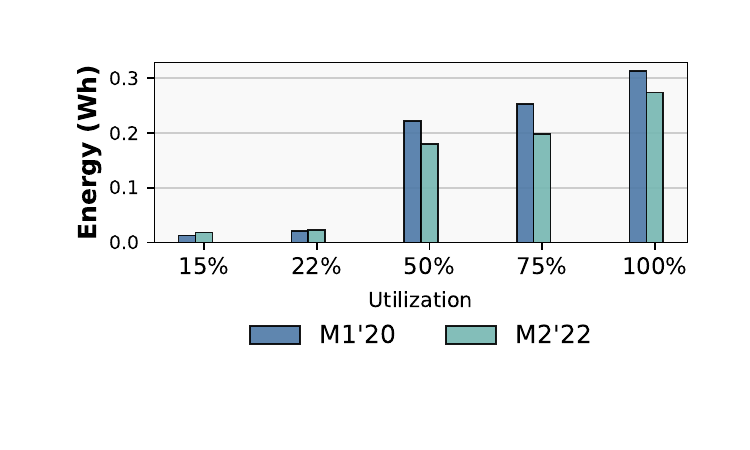}
    \vspace{-0.2in}
    \caption{Energy consumption versus utilization across machines.}
    \label{fig:util}
\end{figure}
{\bf Energy and Utilization.}
We further analyzed how varying utilization levels affect power consumption across different server generations. To do this, we ran a set of workloads designed to achieve different utilization profiles and measured their energy consumption. Figure~\ref{fig:util} shows the energy usage for two machines (from 2020 and 2022), under varying levels of utilization. We observe that at comparable low utilization levels, the 2020 server consumes less energy than the newer 2022 model. However, as utilization increases, the 2022 server becomes more efficient overall, consuming less energy for the same amount of work. This indicates that newer servers may only realize their energy efficiency advantages under high-utilization workloads, and may in fact be less efficient for lightly loaded scenarios. \textit{These findings suggest that under low to moderate utilization, older platforms may remain competitive in lifecycle efficiency when workload deadlines are modest and embodied replacement costs are significant.}

\begin{figure}[t]
    \centering
    \includegraphics[width=2.8in]{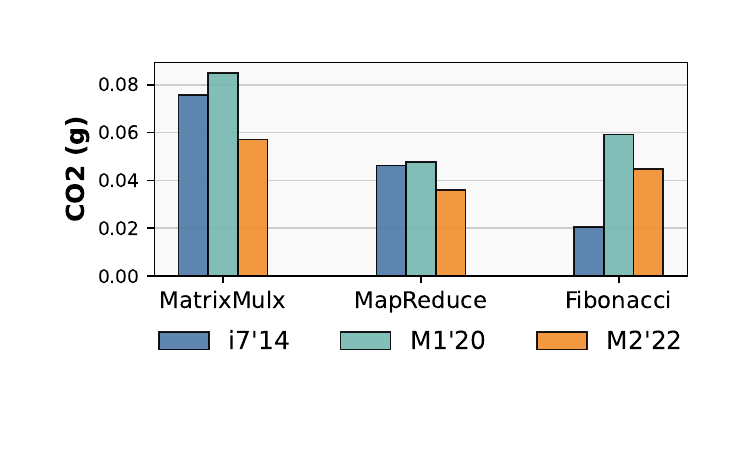}
    \caption{Emissions when servers powered by 100\% renewable energy.}
    \label{fig:carbon}
\end{figure}
{\bf Low-Carbon Electricity Motivation.}
A second motivation for extending hardware lifetime arises from electricity-grid decarbonization. As electricity grids incorporate larger shares of low-carbon generation, operational emissions associated with compute workloads decrease substantially, shifting the relative carbon burden toward embodied hardware emissions. In such settings, already-deployed machines may retain lifecycle advantages even when their operational efficiency is lower.

Figure~\ref{fig:carbon} illustrates lifecycle carbon estimates across hardware generations under low-carbon electricity assumptions, where embodied emissions become a dominant component of total emissions. \textit{These observations suggest that under low-carbon electricity conditions, upgrade decisions become increasingly sensitive to workload execution characteristics, assumed service life, and embodied carbon allocation.}

\section{CarbonSim Design}
\begin{figure}[t]
    \centering
    \includegraphics[width=3.2in]{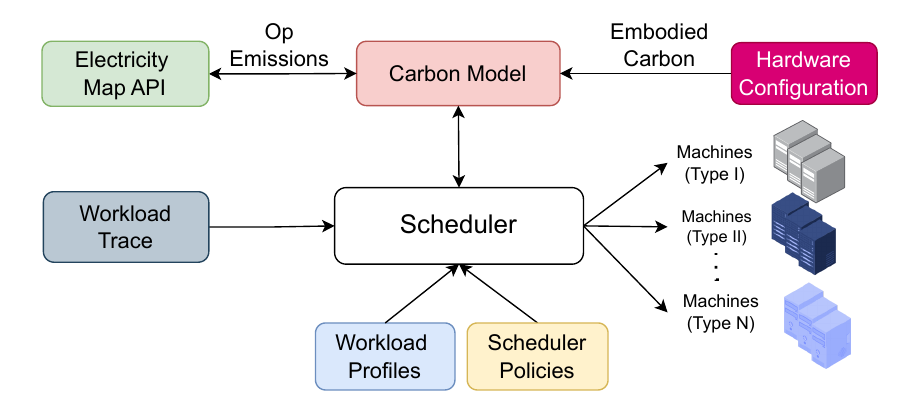}
    \caption{Overall architecture of \texttt{CarbonSim}.}
    \label{fig:architecture}
\end{figure}

Figure~\ref{fig:architecture} presents the architecture of \texttt{CarbonSim}, a lifecycle-aware simulation framework for evaluating carbon and performance tradeoffs associated with hardware upgrade decisions across heterogeneous compute fleets. CarbonSim combines machine-level hardware profiles, workload traces, workload-specific execution models, embodied carbon inventories, scheduling policies, and time-varying electricity carbon intensity to estimate lifecycle emissions under alternative deployment scenarios.
A key feature of \texttt{CarbonSim} is its incorporation of detailed carbon accounting models and workload profile information, enabling the simulation to capture both embodied and operational emissions across hardware generations.

\texttt{CarbonSim} currently models compute-node emissions explicitly and treats facility overhead as an extensible parameter rather than a fixed component. Cooling systems, networking equipment, memory subsystems, storage devices, and accelerator devices are therefore outside the default system boundary, but can be incorporated through configuration using power usage effectiveness (PUE) multipliers or hardware-specific profiles. 

At its core, \texttt{CarbonSim} functions as a generalized container scheduling system that makes placement decisions based on various factors, such as hardware specifications, workload characteristics, and scheduling policies. The simulation uses hardware configurations to set up the cluster and implements job queues for each machine. When a job request arrives, the simulator evaluates the available machines and assigns the container to one based on the selected scheduling policy.  The simulator then updates the resource status, marking it as in use based on the container's resource requirements, such as CPU and memory. It then leverages workload profiling data to estimate the container's execution time and energy consumption. This information is subsequently used to calculate the total emissions.

{\bf Carbon Model.} We compute the overall emissions as a combination of operational and embodied emissions.
\begin{equation*}
    tot\_emissions  = \sum_{t=1}^{T} op\_emissions_t(s) + embodied\_emissions_t(s)
    \label{emission_equation}
\end{equation*}
where $T$ is the simulation duration and $s$ represents the machine. 
Furthermore, the operational emissions is: $op\_emissions_t(s) = carbon\_intensity_t \times energy_t(s) \times T_{exec}(s)$. Here, the energy consumption and job execution time, $T_{exec}(s)$, are derived from the workload profile input data, which includes information about the execution times and energy use of workloads across different machines. Additionally, the carbon intensity information is the forecasts of the carbon intensity of the electricity grid from the Electricity Map API~\cite{elecmap}.

For embodied emissions, we use a hardware configuration file where users can input the total emissions associated with a given machine. \texttt{CarbonSim} supports two accounting approaches for embodied emissions: (a) \textit{fixed} and (b) \textit{frontloading}. In the \textit{fixed model}, the embodied emissions are evenly distributed over the estimated lifespan of the system, assuming that the server's emissions are spread uniformly throughout its operational life. In contrast, the \textit{frontloading model} allocates a larger proportion of the embodied emissions to the initial phase of the server's lifecycle, with emissions decreasing over time. This model reflects the idea that a significant portion of a system's embodied emissions is "front-loaded" during the production, transportation, and installation stages, rather than being distributed evenly across its entire operational lifespan. \texttt{CarbonSim} uses an exponential decay function to model the higher environmental impact of newer servers in their early stages of use.

{\bf Scheduling policies.} 
\texttt{CarbonSim} includes several scheduling strategies to simulate hardware upgrade decisions. We have adapted four existing scheduling policies, though the framework can be extended to include more. Specifically, we implement the following: \textit{Carbon-Aware}, \textit{SLO-Aware}, \textit{Wait-a-While}~\cite{wiesner2021let}, and \textit{Threshold-based Greedy approach}~\cite{souza2023ecovisor}.

In \textit{Carbon-Aware} policy, workloads are assigned to machines that produce the least overall emissions.
The \textit{SLO-Aware} policy allows delaying workloads to minimize carbon emissions, as long as the delay stays within a Service Level Objective defined as a percentage of overhead compared to a baseline configuration.
The \textit{Wait-a-While} policy schedules workloads in time slots when grid carbon intensity is lowest, exploiting hourly fluctuations to reduce emissions.
Finally, the \textit{threshold-based greedy} policy executes the job if the current carbon intensity is below the 30th percentile of the next 24-hour forecast; otherwise, it pauses the job.


{\bf Measurement.} Machine-level power profiles are obtained using platform-specific instrumentation. Intel i7 systems are measured using \textit{Intel RAPL} package energy counters, Apple M1/M2 platforms are sampled using macOS \textit{powermetrics}, and older Core2 systems are measured using an external wall-power meter sampled at 1 Hz. Because energy interfaces differ across generations, all measurements are normalized to workload-only execution intervals after excluding idle background activity. The framework allows server-grade power profiles, which can be provided using the hardware configurations. 

{\bf Implementation.} We implement the simulator in Python. The simulation operates in discrete-event time, modeling time progression in defined steps to ensure an efficient and detailed analysis. \texttt{CarbonSim} is also extensible and allows users to incorporate new scheduling strategies and workload profiles. The simulator allows the user to select the desired scheduling policy and degradation model for experimentation. Additionally, the user is required to provide an Excel sheet containing comprehensive details, including workload profiling for various machines, machine configurations, year of manufacture, embodied emissions, and inventory data specifying the available instances of each machine. To enhance the accuracy of carbon emissions calculations, we integrated the Electricity Map API to fetch real-time carbon intensity data. An active API key is required to access this data.

\section{Results}

\begin{table}[t]
\centering
\small
\begin{tabular}{|p{2cm}|p{5.5cm}|}
\toprule
\textbf{Hardware} (year)  & Intel Core Duo (2009), Intel Core i7-4770 (2013), Intel Core i7-4790 (2014), Apple M1 (2020), Apple M2 (2022) \\
 & \\\hline

\toprule \hline
\textbf{Workload} & Matrix Multiplication, MapReduce, Fibonacci, GetPrime \\\hline

\toprule \hline
\textbf{Location}(Carbon Intensity gCO2/kWh) & Quebec (42), Spain(146), California(242), Queensland(579)
\\\bottomrule
\end{tabular}
\caption{Hardware and workload description.}
\label{tab:hardware_workloads}
\end{table}
We evaluate the carbon savings achieved by utilizing older machines across various scenarios. These platforms include older Intel-based systems and recent Apple and Intel machines chosen to capture broad differences in idle power, execution efficiency, and performance scaling across hardware generations. Table~\ref{tab:hardware_workloads} summarizes the hardware configurations and workloads used in our study. These workloads were selected to capture a diverse set of computational patterns and resource requirements. For each workload, we measured execution time, energy consumption, and CPU utilization to analyze how they perform across different hardware generations.


\subsection{Carbon-Performance Variations}
\begin{figure}[t]
    \centering
    \includegraphics[width=2.8in]{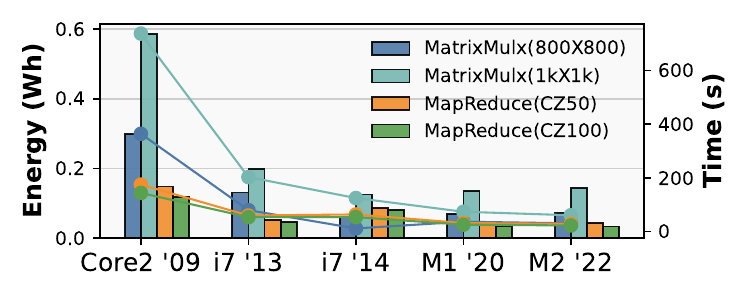}
    \caption{Workload profile of workloads across machines.}
    \label{fig:workload}
\end{figure}

\begin{figure}[t]
    \centering
    \includegraphics[width=2.8in]{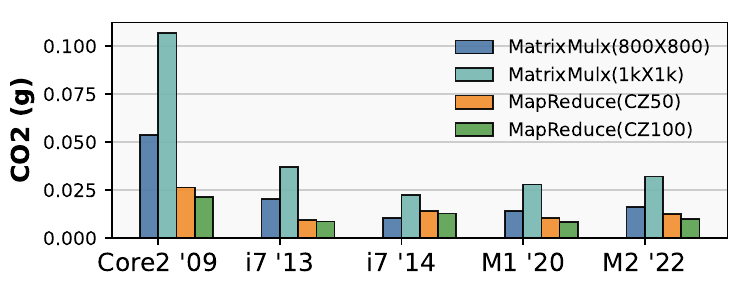}
    \caption{Carbon footprint of workloads across machines.}
    \label{fig:carbon}
\end{figure}
We ran the workloads on different machines and used RAPL (Running Average Power Limit) to measure energy consumption. Figure~\ref{fig:workload} shows the workload profiling results, including execution time and energy consumption across various machine generations. As expected, newer machines demonstrate reduced execution time and energy consumption compared to older machines. However, the overall carbon emissions do not follow the same trend. Figure~\ref{fig:carbon} shows that carbon emissions are nearly comparable between older and newer machines. Despite the longer execution times on older machines, when accounting for embodied emissions, the total emissions are often lower or comparable to those of newer machines. This outcome highlights the significance of embodied emissions in the overall carbon footprint.
We believe this trend is likely to persist, as energy-efficiency improvements in hardware are becoming less significant with each new generation. This suggests that extending the use of older machines, particularly in scenarios where embodied emissions are a substantial factor, could be a viable strategy for reducing the overall carbon footprint.

\subsection{Renewable Energy Impact}


\begin{figure}[t]
    \centering
    \includegraphics[width=2.8in]{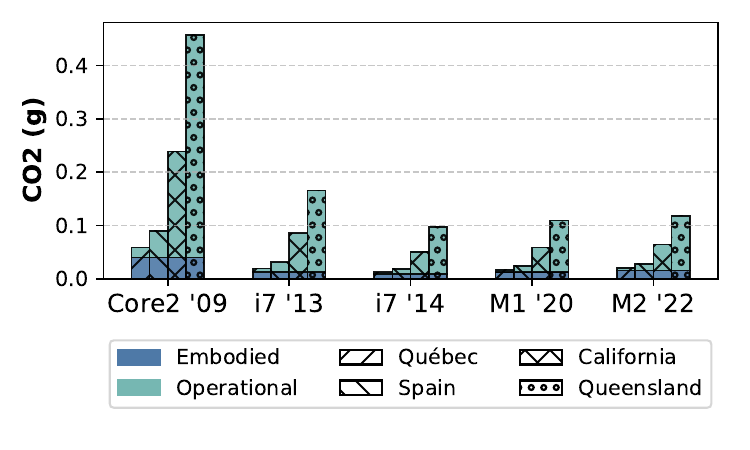}
    \caption{Carbon footprint comparison across locations.}
    \label{fig:location}
\end{figure}

We also investigate how different carbon intensities affect overall emissions during workload execution. For this experiment, we ran the Matrix Multiplication workload on different machines while varying the geographic location, each with distinct carbon intensity levels. Figure~\ref{fig:location} illustrates the impact of location-specific energy grid composition on overall emissions. Our analysis reveals that in locations with a high share of renewable energy (e.g., Quebec), embodied emissions dominate the overall carbon footprint~\cite{radovanovic2022carbon}. In such cases, continuing to use older machines with already-incurred embodied emissions can be more environmentally favorable. 
Conversely, in locations with high carbon intensity, operational emissions become the primary contributor to the overall footprint. Under these conditions, upgrading to newer, energy-efficient machines can yield significant emission reductions. This suggests that the decision to upgrade hardware should be informed by the carbon intensity of the location. For regions transitioning to cleaner energy grids, extending the lifespan of existing hardware may provide greater environmental benefits. However, in areas with carbon-intensive energy grids, investing in newer, more efficient hardware may be more advantageous for reducing overall emissions. 




\subsection{Embodied Accounting Model}
\begin{figure}[t]
    \centering
    \includegraphics[width=2.8in]{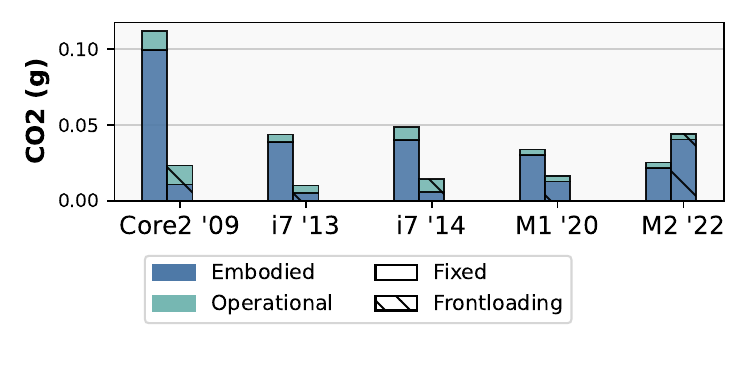}
     \caption{Carbon footprint of embodied accounting model.}
    \label{fig:account}
\end{figure}

We also evaluate the impact of the embodied emissions accounting model on overall emissions. For this experiment, we ran the MapReduce workload across different machines and set the experiment year to 2022 to analyze how the frontloading model influences total emissions, using embodied emissions values obtained from the manufacturers' official websites. Figure~\ref{fig:account} compares the total emissions calculated using the fixed and frontloading models.

We observe that under the fixed emissions model, the embodied emissions are attributed evenly over the server's entire lifespan, leading to a consistent contribution to overall emissions. In contrast, the frontloading model emphasizes the early stages of the server's lifecycle, resulting in higher embodied emissions during the initial years and significantly lower emissions in later years. Thus, for newer machines, the frontloading model results in higher overall emissions due to the significant emphasis on embodied emissions during the server's early years. Interestingly, in both fixed and frontloading models, newer machines can have higher total emissions than older machines when accounting for both embodied and operational contributions. This suggests the importance of carefully considering embodied emissions in lifecycle assessments and hardware upgrade decisions.

\subsection{Scheduling Policies Effect}
\begin{figure}[t]
    \centering
    \includegraphics[width=2.8in]{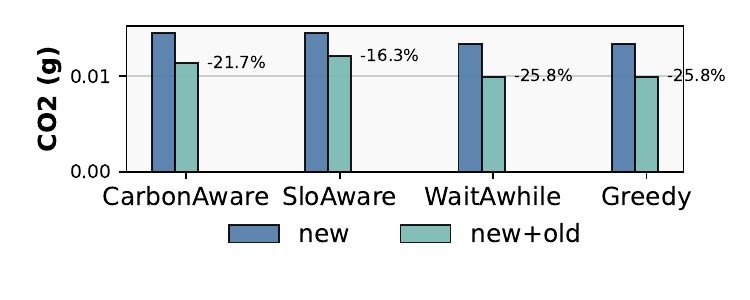}
    \caption{Carbon emissions of various scheduling policies.}
    \label{fig:policy}
\end{figure}

\begin{figure}[t]
    \centering
    \includegraphics[width=2.8in]{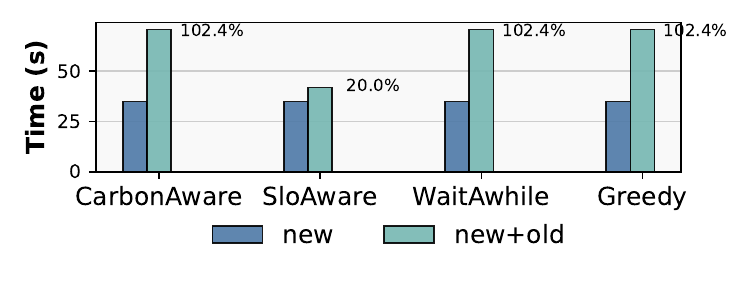}
    \caption{Execution time across various scheduling policies.}
    \label{fig:performance}
\end{figure}
We evaluate the impact of different scheduling policies across various cluster configurations. For this analysis, we consider two clusters: one consisting of 10 new machines, and another composed of a mix of 10 machines, with two machines from each generation spanning 2009 to 2024. We also assume varying workload traces to simulate diverse operating conditions. Figure~\ref{fig:policy} presents the overall carbon footprint across different scheduling policies. The results show that clusters with a mix of old and new machines achieve a lower carbon footprint compared to clusters with only new machines. However, this reduction in emissions comes with a trade-off --- a noticeable degradation in performance, as reflected by increased execution times shown in Figure~\ref{fig:performance}. For some policies, the execution time doubles when using mixed-generation clusters.
These results suggest that while it is possible to significantly reduce overall emissions by incorporating older machines into the cluster, designing scheduling policies that carefully consider both carbon efficiency and performance is essential. Such policies can ensure that environmental benefits are achieved without imposing excessive performance penalties.
We also experimented with clusters of varying sizes (not shown in the figure) and observed a similar trend across all configurations: a consistent reduction in overall emissions when incorporating older machines into the cluster. However, this reduction in emissions was accompanied by a noticeable decline in performance, reflecting the tradeoff in using mixed clusters.

\section{Discussion and Future Work}
Recent studies have extensively focused on managing and optimizing energy efficiency and operational carbon emissions~\cite{elgamal2023carbon, wang2024designing, wiesner2024vessim}. \texttt{CarbonSim} builds upon prior work by incorporating embodied carbon into scheduling decisions, alongside operational emissions, to better understand the trade-offs of hardware upgrades from an environmental perspective.
A key limitation is that our system does not account for other costs associated with older machines, such as maintenance and reliability issues. However, extending machine lifetime can still offer environmental benefits, particularly in regions where the energy grid is transitioning to renewable sources. Another limitation is the reliance on workload profiling information, which may not fully capture workload complexity. Nevertheless, this provides a reasonable approximation, as VMs/containers can be allocated dedicated resources, helping mitigate performance trade-offs. In future work, we plan to approximate profiles using benchmarks and to implement the framework on container orchestration platforms such as Kubernetes.

\section{Conclusion}
We presented CarbonSim, a lifecycle-aware simulation framework for evaluating carbon tradeoffs in hardware upgrade decisions by combining workload execution profiles, machine-level power characteristics, embodied carbon inventories, scheduling policies, and time-varying grid carbon intensity. Our results show that newer hardware does not always minimize carbon emissions — performance and energy-efficiency gains may not offset the embodied carbon cost of replacement, particularly for lightly loaded workloads and cleaner electricity grids. In such cases, extending existing hardware can reduce lifecycle emissions. Overall, hardware refresh decisions should be evaluated using lifecycle-aware models and must be workload-aware, location-aware, and lifecycle-aware.

\begin{acks}
This work is supported in part by NSF grant \#2324873 and Department of Energy \#DECR0000041. The authors would also like to acknowledge the Mascaro Center for Sustainable Innovation at the University of Pittsburgh for its support.
\end{acks}

\bibliographystyle{ACM-Reference-Format}
\bibliography{paper}


\end{document}